Ozone depletion-induced climate change following a 50 pc supernova

Brian C. Thomas and Cody L. Ratterman

Washburn University, Department of Physics and Astronomy
1700 SW College Ave., Topeka, KS USA 66621
brian.thomas@washburn.edu

Abstract:
Ozone in Earth's atmosphere is known to have a radiative forcing effect on climate. Motivated by geochemical evidence for one or more nearby supernovae about 2.6 million years ago, we have investigated the question of whether a supernova at about 50 pc could cause a change in Earth's climate through its impact on atmospheric ozone concentrations. We used the "Planet Simulator" (PlaSim) intermediate-complexity climate model with prescribed ozone profiles taken from existing atmospheric chemistry modeling. We found that the effect on globally averaged surface temperature is small, but localized changes are larger and differences in atmospheric circulation and precipitation patterns could have regional impacts.

## I. Introduction

Discovery of live $^{60}Fe$ in ocean sediments at around 2.6 million years ago (with a weaker signal about 8 million years ago) has established the relatively nearby explosion of one or more core-collapse supernovae (SNe) within a range of 50-100 pc [1–7]. The Pliocene-Pleistocene boundary occurs at about 2.6 million years ago, and is characterized by a general cooling of the climate, extinctions, and an increase in wildfires [8]. Supernovae produce electromagnetic emission (UV, X-ray, gamma-ray) and also accelerate nuclei to high energy, producing cosmic rays (CRs).

A robust feature of nearby supernovae, and astrophysical ionizing radiation events in general, is depletion of stratospheric ozone ($O_3$) in Earth's atmosphere [9–11]. This is important from the perspective of life on Earth since $O_3$ strongly absorbs biologically-damaging solar UVB radiation (wavelength 280-315 nm). Recent modeling studies have examined the effects of supernovae at 100 pc and 50 pc on Earth's atmosphere biosphere [12–14]. At 100 pc, effects were found to be minor [13]. At 50 pc, changes in UV irradiance and ground-level muon dose could be significant for some organisms [12,14].

Another potential impact of changes in atmospheric $O_3$ concentrations not previously explored is the effect on Earth's climate. This possibility for the case of a nearby supernova was first explored by [15]. Besides strong absorption in the UV, $O_3$ also has weaker absorption bands in visible and infrared (IR) wavelengths and therefore plays a role in the radiative forcing energy budget. In addition, absorption of UV by $O_3$ heats the stratosphere and changes in concentration can affect atmospheric heating and thereby transport, leading to regional or even global changes in winds, precipitation, and surface temperatures [16–18]. Such changes have been observed across the Southern Hemisphere due to recent anthropogenic $O_3$ depletion over Antarctica [17,19].

Previous radiative transfer modeling for the 50 pc case found small change in surface-level visible light irradiance due to changes in $O_3$ [14], but that work was not able to evaluate the potential for changes in Earth's climate. Here, we investigate this possibility using a global climate model of intermediate complexity with $O_3$ profiles from previous modeling [12].

## II. The Supernova Case

Supernovae are astrophysical explosions that signal the end-point of the life of some types of stars. There are a number of different types of supernovae, classified primarily by observational features such as spectra. Some supernovae are the result of "core-collapse" of a massive (more than about six solar masses) star after it has run out of fusion fuel. Another type occurs when a stellar remnant called a white dwarf gathers mass from a companion star in a binary system and passes the limit of stability, resulting in the explosion of the white dwarf. All supernova types are characterized by a very rapid increase in visible light brightness. Some types have also been observed to emit high-energy radiation in the form of x-rays and gamma-rays. In addition, supernova explosions (at least some types) are the most likely source for the observed flux of high-energy nuclei known as "cosmic rays" (CRs). Cosmic rays and high-energy photons affect Earth's atmospheric chemistry by producing ionizations, dissociations, dissociative ionizations and excitations of atmospheric constituents, primarily $N_2$. The details depend mostly upon the energy spectrum of the incident radiation. In general, these changes lead to production of odd-nitrogen compounds (most importantly NO, $NO_2$) which participate in catalytic cycles that destroy ozone. More details of the types of supernovae, their effects, and rates at which they may have significant impact on Earth can be found in Ref. [11].

We consider here the case of a type IIP supernova (the most likely type associated with the geochemical evidence [20]) at about 50 pc. While SNe produce high-energy photons, this particular type at a distance of 50 pc would not deliver significant flux in this form. On the other hand, as described in [12,13], Earth would receive a significant flux of accelerated protons (and heavier nuclei, together known as cosmic rays) exceeding the normal background for as long as several hundred thousand years. Here we adopt the most likely (and conservative) cosmic ray flux case examined previously ("Case B" in Ref. [12]). The maximum effect on stratospheric $O_3$ in that case was found for the cosmic ray flux at around 300 years after arrival of the first photons from the SN. The flux can be considered to be roughly constant over periods of a few hundred years.

## III. Atmospheric Chemistry Modeling and Ozone Changes

Atmospheric chemistry modeling (detailed in [12,14]) was performed using the Goddard Space Flight Center (GSFC) 2D (latitude and altitude) model. The model and methods used for simulating the effect of supernova-induced ionization have been extensively described elsewhere [10,12,14]. Here we summarize the main features of the model and methods used in those studies. The GSFC model has 18 equal bands in latitude from pole to pole, 58 evenly spaced logarithmic pressure levels (from the ground to approximately 116 km), sixty-five chemical species, winds, small-scale mixing, solar cycle variations. The model also includes heterogeneous processes, most importantly reactions mediated by solid nitric acid trihydrate (NAT) particles in polar stratospheric clouds, which are especially important in Antarctic ozone depletion. The model has been used extensively for studies of atmospheric chemistry changes under a variety of scenarios, including gamma-ray bursts [10], supernovae [9,12,13], and solar proton events [21,22]. For the case of a supernova, latitude-dependent vertical profiles of ionization caused by a spectrum of cosmic radiation are pre-computed and then used as a source of odd-nitrogen and odd-hydrogen compounds in the GSFC model. The model then runs under these conditions, simulating the chemical response throughout latitude and altitude, producing 2D profiles of chemical constituents including $O_3$.

Since the CR flux over a few hundred years can be treated as steady-state, the ionization profiles were treated as time-independent and the model was run until equilibrium was achieved (about 10 years). In addition, since the CR particles travel from the SN to Earth in a diffusive way, the received CR flux is isotropic and the ionization is globally uniform. For the SN case described above Refs. [12,14] found $O_3$ depletion of about 25% globally averaged, up to about 40% over a given latitude.

## IV. Climate Model Setup

In order to investigate possible climate effects of previously modeled changes in $O_3$ we used the "Planet Simulator" (PlaSim) model, described in [23–26] (freely available at https://www.mi.uni-hamburg.de/en/arbeitsgruppen/theoretische-meteorologie/modelle/plasim.html). The model is a coupled ocean-sea ice-atmosphere general circulation model. The model includes radiative effects of $O_3$ and has been utilized for similar studies of climate effects by changes in $O_3$ [25,26]. We used the model with its standard modern-Earth configuration (solar irradiance, orography, and land/sea mask), with default initial conditions, 10 vertical levels with the top level at 40 hPa, and T21 horizontal resolution (32 latitudes, 64 longitudes). For all results described below we ran the model for 100 model years.

While our general motivation is the case of a nearby supernova around 2.6 million years ago, this work is not an attempt to model effects under climate conditions at that time. Rather, our goal is to determine if any significant climate impact might be expected in general, so we have taken the most well-tested configuration of the model as our starting point. We ran the model with interactive (rather than prescribed or static) mixed-layer ocean temperatures and sea ice, and a fixed $CO_2$ concentration of 360 ppm.

It is known that $CO_2$ concentration has varied widely in the past. To investigate whether our results depend on the background $CO_2$ concentration we performed additional simulation runs with 180 ppm and 1000 pm, which bracket the most likely extreme values over the past several hundred million years [27]. The results, comparing runs with and without supernova influence, are very similar for every $CO_2$ concentration case. Variation in $CO_2$ does not have a significant impact on our conclusions.

We validated our implementation of the model by reproducing results in [25,26] which studied the climate effect of removing $O_3$ entirely. We found good agreement for runs with similar setup.

## V. Prescribed $O_3$ Profiles

In the case of an ionizing radiation event, $O_3$ is typically depleted at higher altitudes but increased some at lower altitudes (due primarily to an effect known as “self-healing” with some contribution from increased odd-nitrogen mediated “photosmog” reactions) [16,28,29]. The climate effects of $O_3$ depend on its vertical distribution; it is most effective as a greenhouse gas in the troposphere and changes in concentration around the tropopause are most significant for changes in surface temperature [16,30]. Concentration and distribution of $O_3$ is also highly dependent on latitude and season. Therefore, the prescribed $O_3$ profile is an important aspect in climate modeling studies.

The PlaSim model by default utilizes an idealized (analytic) $O_3$ distribution that varies annually and with latitude. This idealized distribution fits well with observed mid-latitude winter $O_3$ distribution [25]. However, for this study we wished to identify the climate effect of an $O_3$ distribution specific to our SN case, as computed by a separate atmospheric chemistry model (the GSFC model, described above). In order to make the closest comparison we needed to use a control $O_3$ distribution from the GSFC model, without input of SN-induced changes. We therefore used a monthly climatology of latitude-dependent $O_3$ profiles taken from the previously completed atmospheric chemistry modeling for both cases.

In order to utilize $O_3$ profiles generated by the GSFC model with PlaSim, we interpolated $O_3$ profiles from the vertical (pressure) and horizontal (latitude) grid used by the GSFC model to that used by PlaSim. The highest vertical level simulated by PlaSim sits at 40 hPa (about 26 km altitude) while the GSFC model extends much higher (to approximately 90 km); the GSFC $O_3$ profiles were therefore truncated at the top-most PlaSim level. The interpolated profiles were then read into PlaSim in place of the default idealized $O_3$ distribution. We used one latitude-dependent monthly $O_3$ profile climatology for a “normal atmosphere” (without SN ionization input) run of the GSFC model, which serves as our control, and another set from a run for the SN case described above. In Figure 1 we show the annual average latitude-dependent $O_3$ profile given by the default idealized distribution (panel A) and the corresponding profiles for the normal GSFC run (panel B, our “custom” control case) and the GSFC run with SN ionization input (panel C). Figure 2 shows the pointwise percent difference between annual average $O_3$ profiles in the default distribution and our “custom” prescribed control distribution. Figure 3 shows the same kind of comparison for the $O_3$ distribution in the SN case versus our custom control.

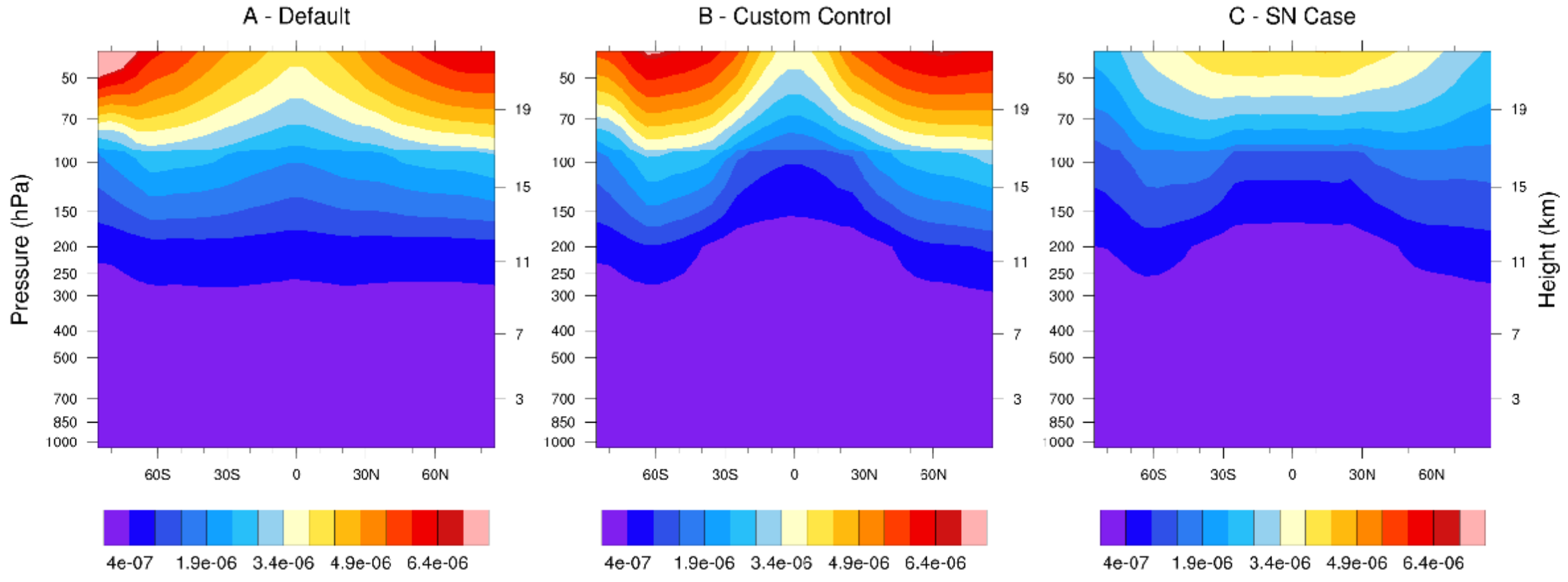

Figure 1: Annual average $O_3$ distribution profiles, in units of kg/kg mixing ratio, for A – the PlaSim “Default” idealized case; B – our “Custom Control” distribution for a normal atmosphere run of the GSFC chemistry model; and C – the “SN case” distribution from the GSFC model run with supernova ionization input. All distributions are uniform in longitude.

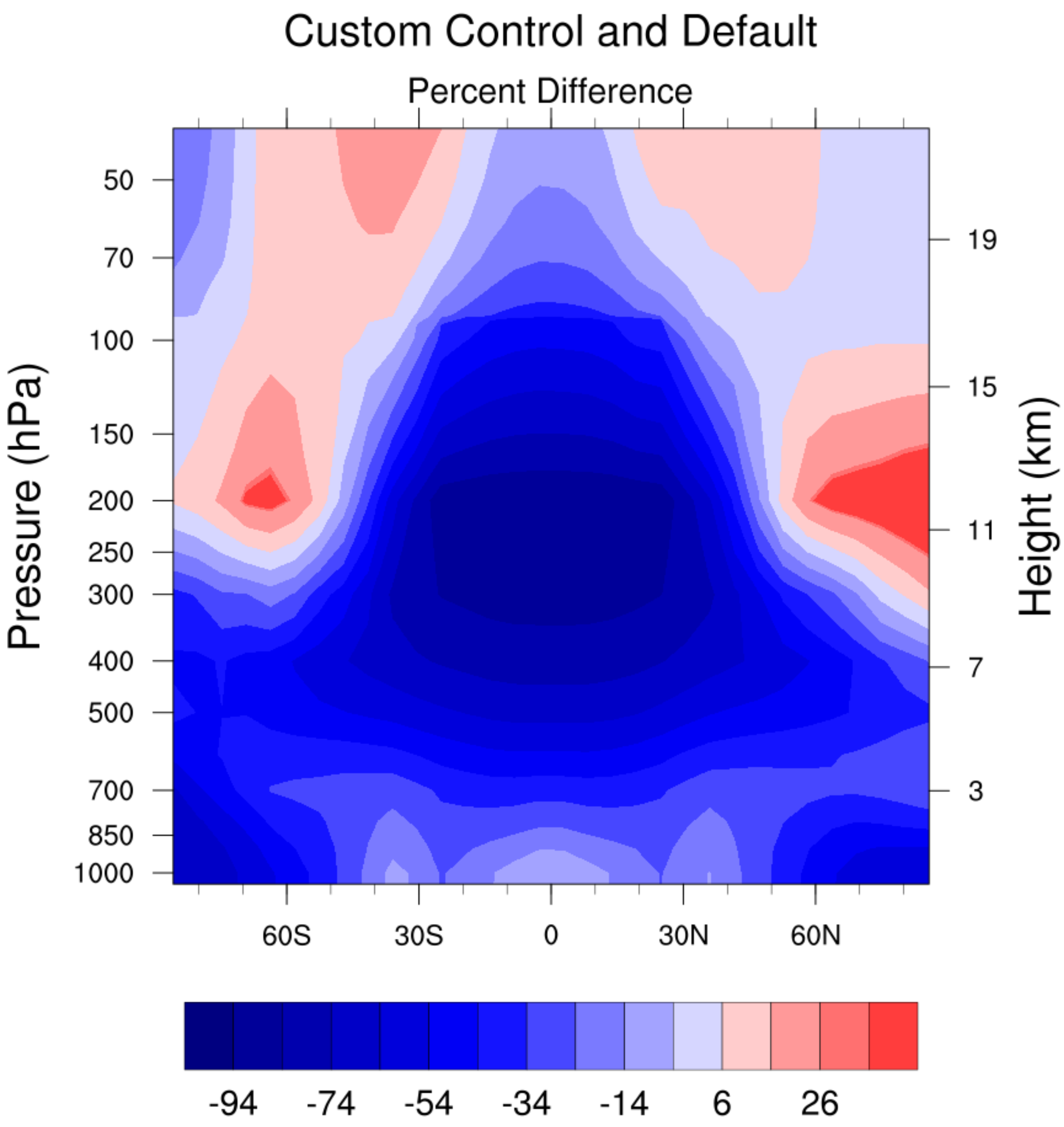

Figure 2: The pointwise percent difference (annually averaged) in $O_3$ concentration between the "Default" PlaSim idealized distribution and our "Custom" control $O_3$ distribution from a normal atmosphere run of the GSFC chemistry model. Positive/negative numbers indicate higher/lower $O_3$ concentration in our "Custom" control case.

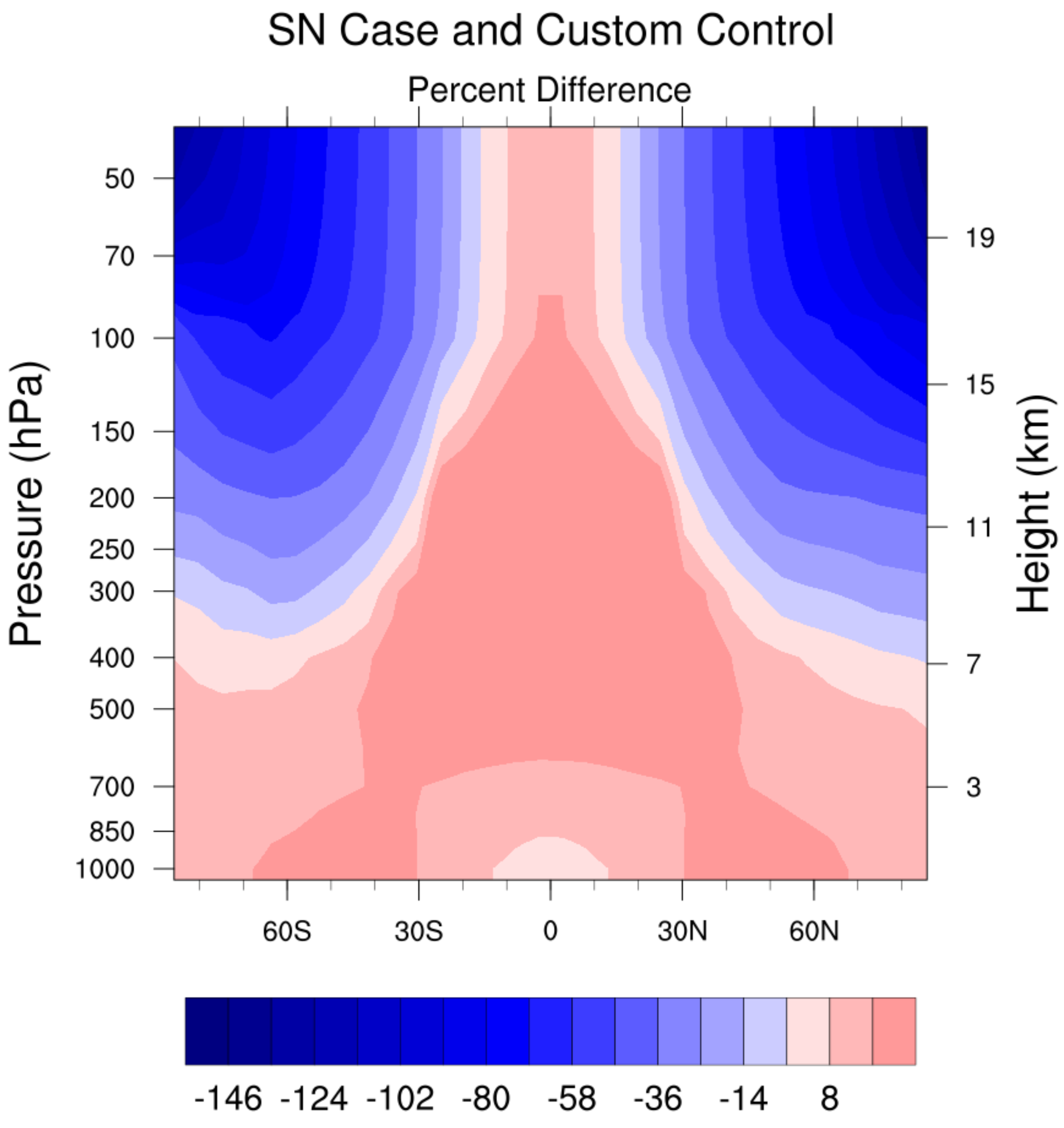

Figure 3: The pointwise percent difference (annually averaged) in $O_3$ concentration between our "Custom" control $O_3$ distribution from a normal atmosphere run of the GSFC chemistry model and the distribution from the GSFC model run with SN ionization input. Positive/negative numbers indicate higher/lower $O_3$ concentration in the SN case.

Figure 4 shows the difference in annually averaged, globally averaged surface temperature between a PlaSim run with our prescribed control $O_3$ profile and a run with the default $O_3$ profile. Both cases ran for 100 model years. After coming to equilibrium, the globally averaged surface temperature with our “custom” control $O_3$ profile is about 2 K lower than with the default $O_3$ distribution. This difference makes sense when one compares the $O_3$ distributions (Figure 1 panels A and B; see also Figure 2). The idealized (default) PlaSim distribution includes much higher $O_3$ around the equatorial and mid-latitude upper troposphere (shown as negative values in Figure 2), where it is most effective as a greenhouse gas. Therefore, with smaller upper troposphere $O_3$ concentration in our prescribed control distribution we would expect a lower surface temperature, as seen in Figure 4.

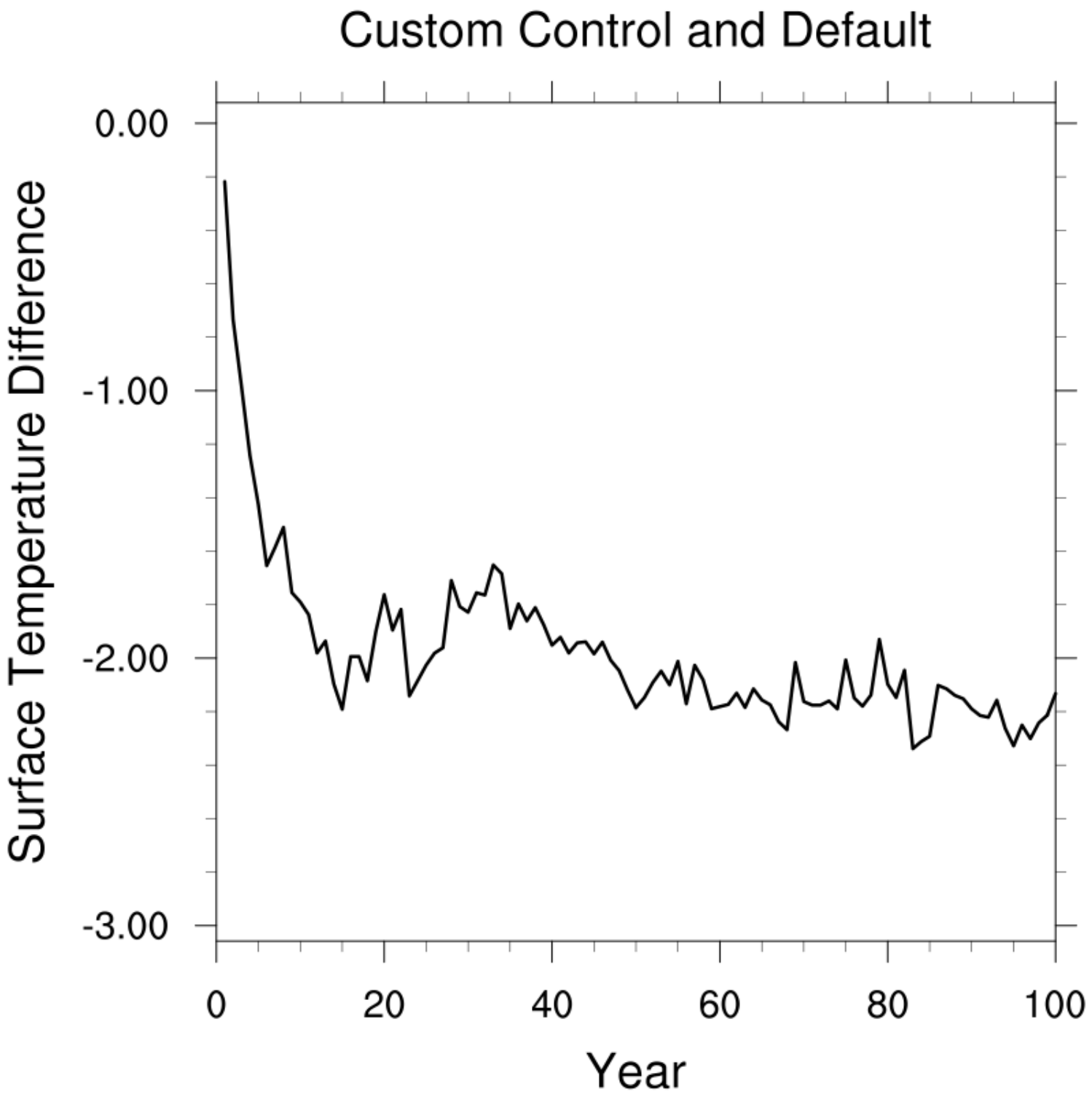

Figure 4: Annually averaged difference (K) between globally averaged surface temperature in the “Default” PlaSim idealized $O_3$ distribution case and the case using our “Custom” control $O_3$ distribution from a normal atmosphere run of the GSFC chemistry model.

## VI. SN-Control Comparison Runs

In order to evaluate the climate impact of $O_3$ changes caused by the nearby supernova case described above, we performed runs of PlaSim with prescribed $O_3$ profiles for our "custom control" and SN cases. Since PlaSim includes a random seed to introduce a noise factor, we performed a set of 10 runs for each case. Figure 5 shows the annually averaged, globally averaged surface temperature over the full 100 years of each run, for all 20 runs, as well as the ensemble mean of each set of 10 for both cases. Two features are apparent. First, the difference in the ensemble mean globally averaged surface temperature is small, only about 0.2 K. Second, this difference is slightly larger than the spread of results. Hence, we conclude that there is only a small difference in globally averaged surface temperature under $O_3$ profile conditions modeled for this SN case. The small difference is likely due to the depletion of $O_3$ being mainly limited to the upper atmospheric levels, even though the total column density depletion is fairly large [12,14].

The slightly higher surface temperature in the SN case can at least partially be attributed to where the $O_3$ changes occur in altitude. Depletion is limited to the higher altitudes, with a small increase in concentration at lower altitudes (where $O_3$ acts most efficiently as a greenhouse gas). Figure 6 shows the zonally averaged profile difference in temperature, again averaged over the ensemble members and time averaged over the last 30 years of the runs. Lower temperatures in the upper levels are due to less heating from $O_3$ absorption of solar UV.

While the global average temperature change is not large, regional effects may still be of interest. In Figure 7 we show the difference between ensemble average surface temperatures (SN vs control) across the globe, averaged over the last 30 years of the runs. Much larger regional differences are seen here (as compared to the global average), primarily in the Polar regions, especially in the Arctic. This fits with results in Ref. [16] which examined a case similar to ours, though with smaller differences in $O_3$ concentration.

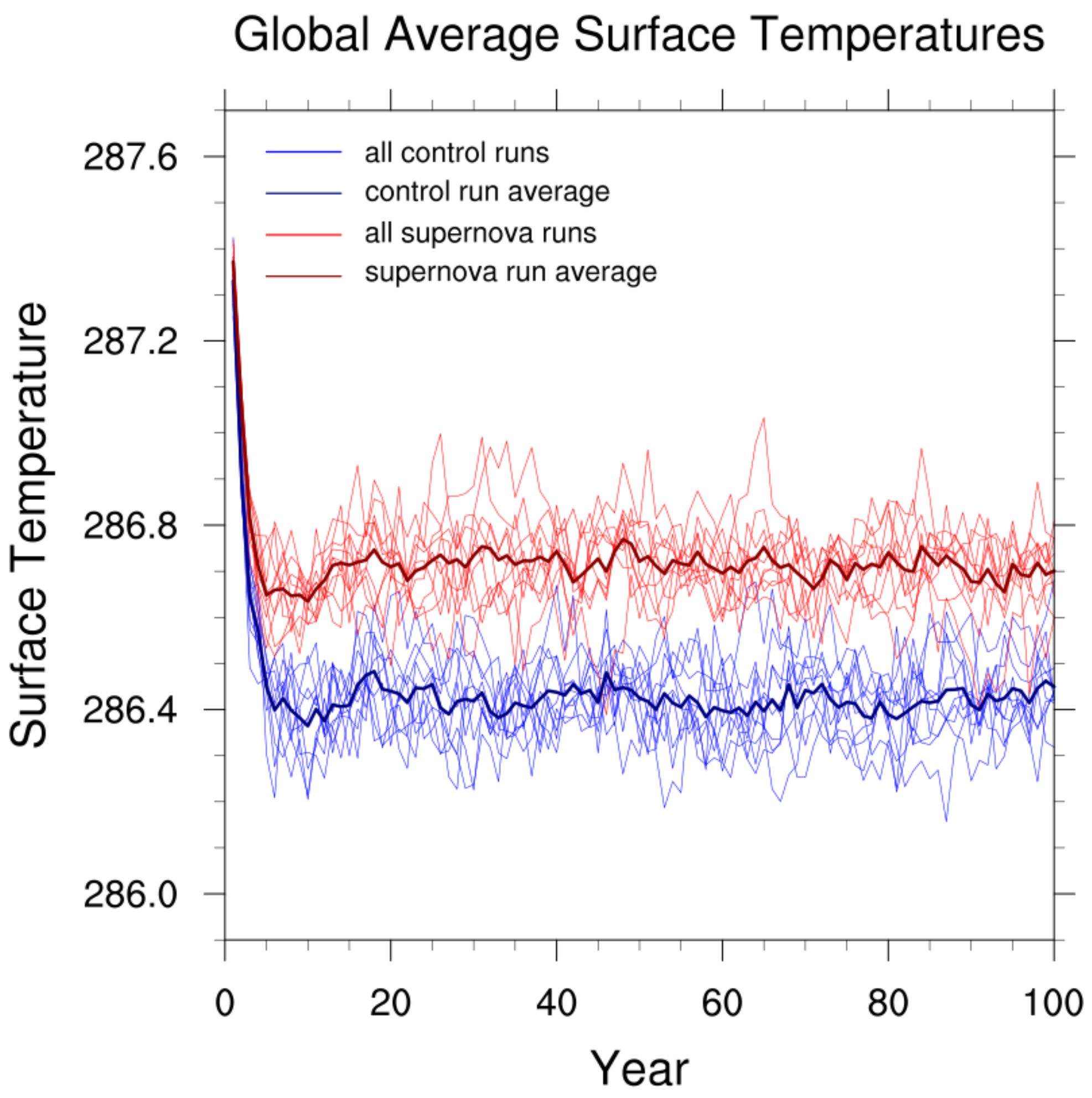

Figure 5: Annually averaged, global average surface temperature (K) for runs with our control (normal atmosphere) $O_3$ distribution (red lines) and runs with our SN case $O_3$ distribution (blue lines). Thin lines show individual runs within the 10-member ensemble for each $O_3$ case; thick lines show the ensemble mean for each case.

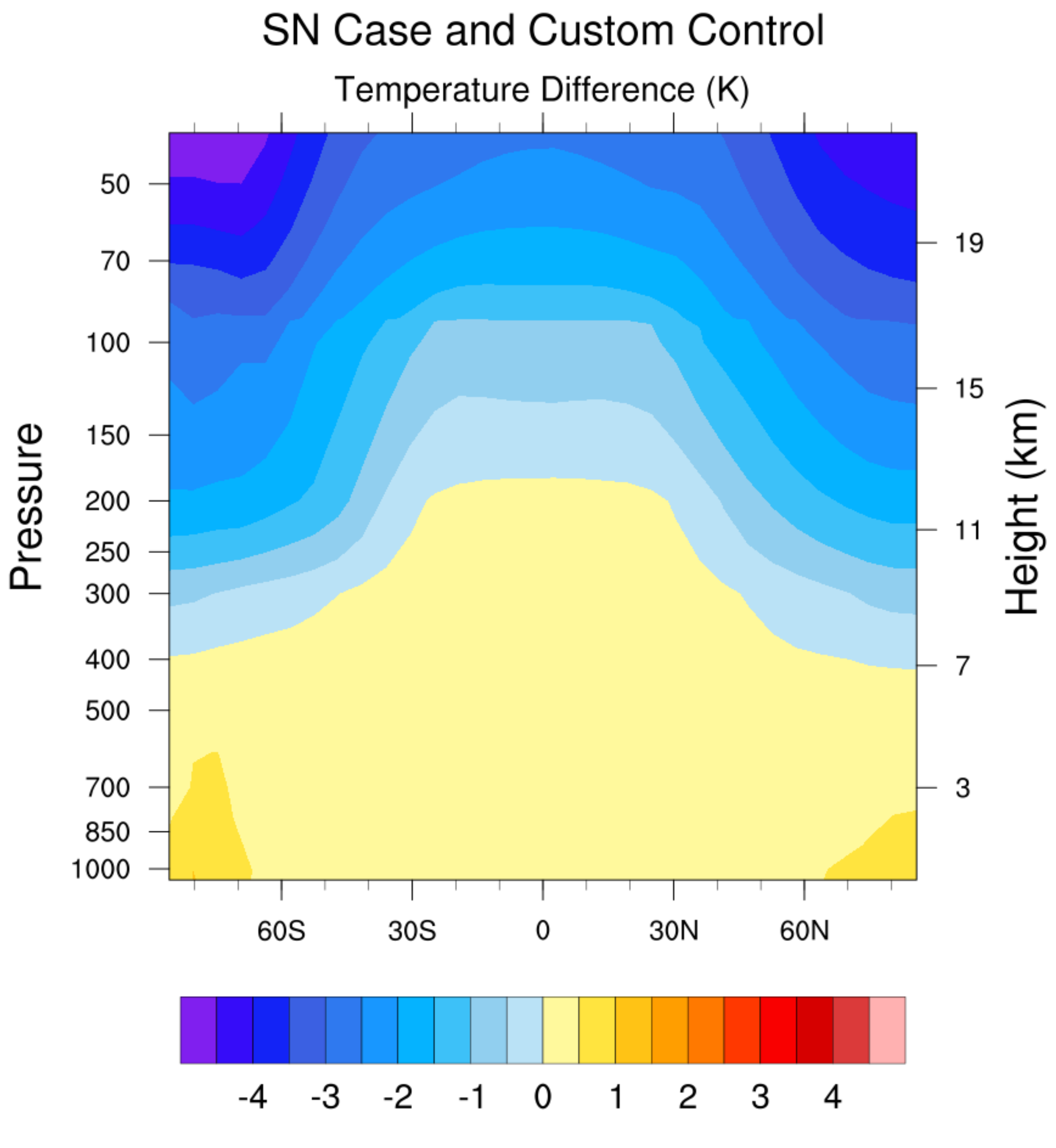

Figure 6: Zonally averaged difference (K) in the ensemble mean temperature profiles for the control and SN cases, averaged over the last 30 years of the runs. Positive/negative numbers indicate higher/lower temperature in the SN case.

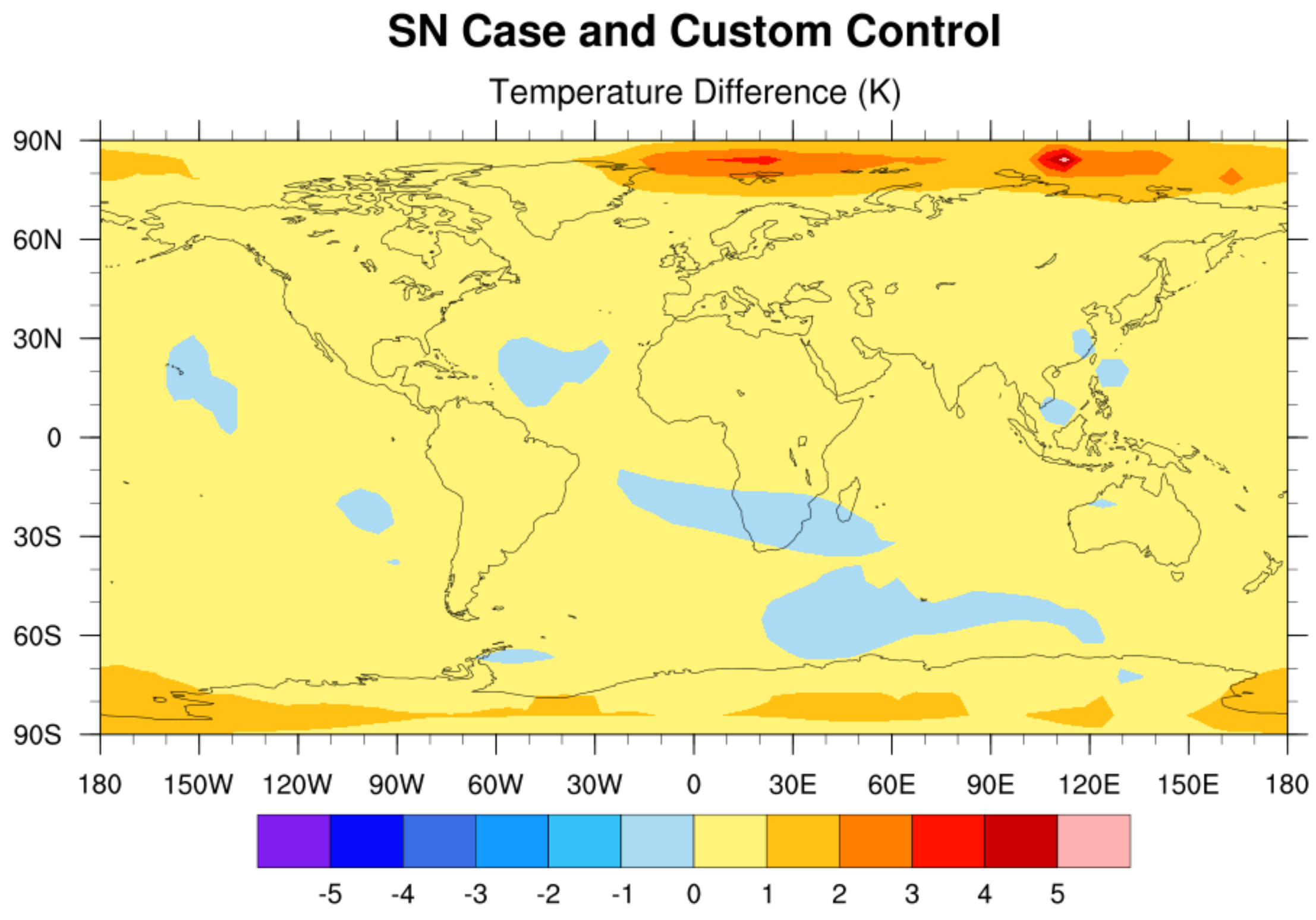

Figure 7: Difference (K) in the ensemble mean surface temperature for the control and SN cases, averaged over the last 30 years of the runs. Positive/negative numbers indicate higher/lower temperature in the SN case.

Regional changes in temperature (and other climate variables) can be tied to changes in atmospheric transport following changes in heating rates due to $O_3$ profile changes. In Figure 6 we see that temperatures are decreased everywhere above the upper troposphere and increased slightly at lower altitudes. Stratospheric cooling due to recent, anthropogenic $O_3$ depletion (primary over Antarctica) has been identified in other studies as a cause of changes in the strength and location of the polar and midlatitude (subtropical) jet streams and corresponding changes in the positioning of the Hadley, Ferrel, and Polar cells [17,19,31,32].

Such changes in the transport features of Earth's atmosphere have a number of consequences, in particular a latitude change of high/low precipitation zones. For instance, a poleward shift of the mid-latitude dry-zone and increased austral summer subtropical precipitation has been reported in observational and modeling studies [17,31,32]. Changes in precipitation has subsequent ecological consequences. Ecosystems in a region accustomed to high rainfall will be negatively impacted by reduced precipitation, and the inverse may be true for areas that are normally dry.

Figure 8 illustrates the effect on atmospheric transport. Each panel shows the difference in zonal (eastward) wind speed between the SN and control cases, at a specific month (January, April, July and October), averaged over the last 30 years of the runs (for the ensemble average). We have chosen these months to highlight the seasonal dependence of the circulation changes, which is linked to the seasonal change in stratospheric heating in the polar regions. January shows a stronger effect in the Southern hemisphere, while July shows a stronger effect in the North, with April and October showing a more symmetric pattern. In the summer-fall periods for each hemisphere the (eastward) polar jet, normally centered around 50° latitude, is increased in strength (positive values) and shifts poleward. Similar trends have been observed and modeled in the Southern hemisphere due to recent anthropogenic ozone depletion over Antarctica [31,32]. In our SN case, $O_3$ depletion is more uniformly distributed in latitude and is high in both North and South polar regions, so we would expect to see such shifts in both hemispheres (in alternating seasons), as shown here.

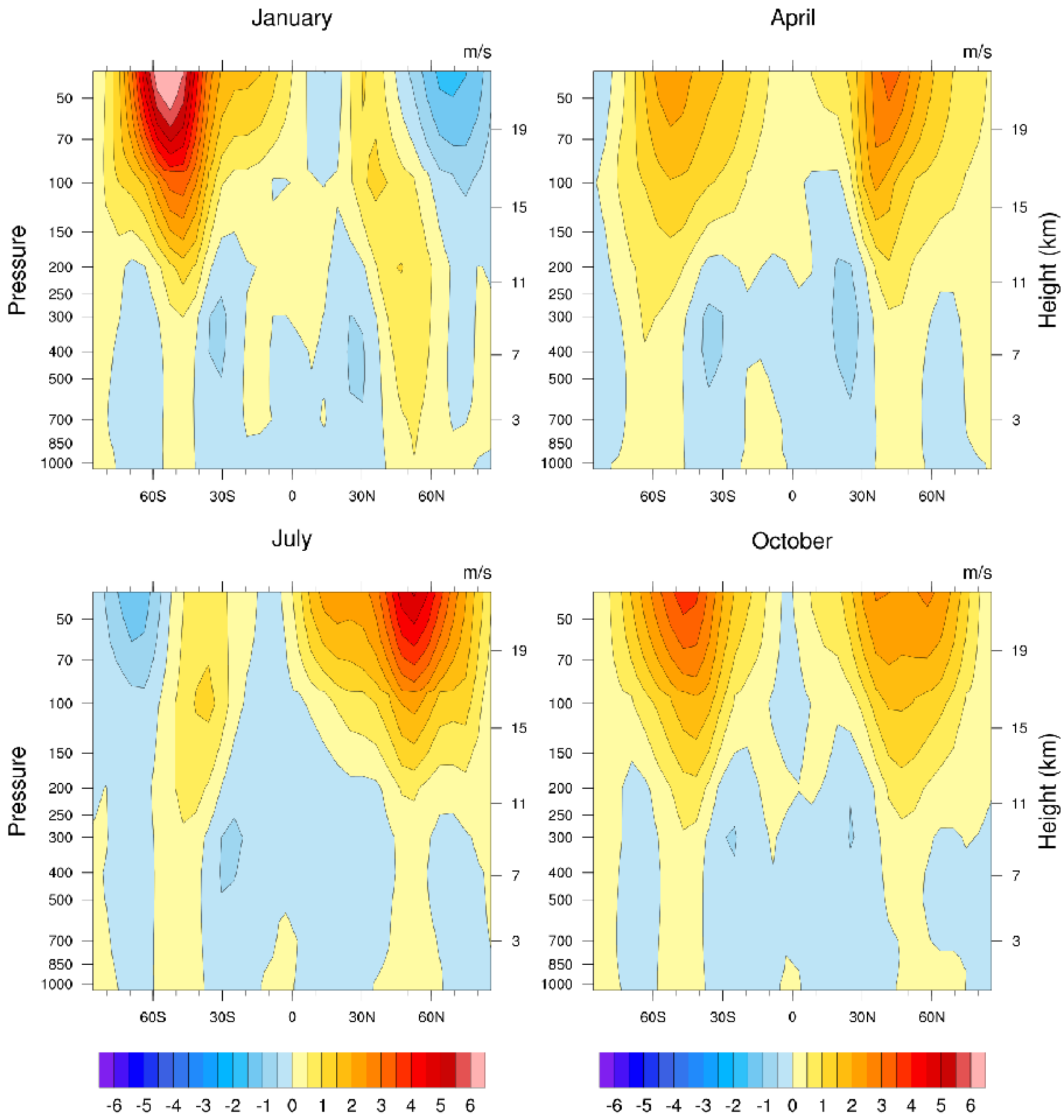

Figure 8: Difference (m/s) in zonal (eastward) wind speed between the SN and control cases, in January, April, July and October, averaged over the last 30 years of the runs (for the ensemble average). Positive/negative numbers indicate higher/lower wind speed in the SN case.

Finally, in Figure 9 we show global changes in precipitation (as percent difference, comparing the SN case to our control case), again averaged over the ensemble and over the last 30 years of the runs.  Here we see reduced precipitation over much of the tropical and midlatitude Southern hemisphere, with a general increase South of about 50° South latitude.  In the Northern hemisphere there is a general increase in precipitation in much of the tropics, with some reduction in the midlatitudes and again a general increase North of about 50° North latitude.  These changes are consistent with shifts in the transport cells that determine zones of consistently higher or lower precipitation.

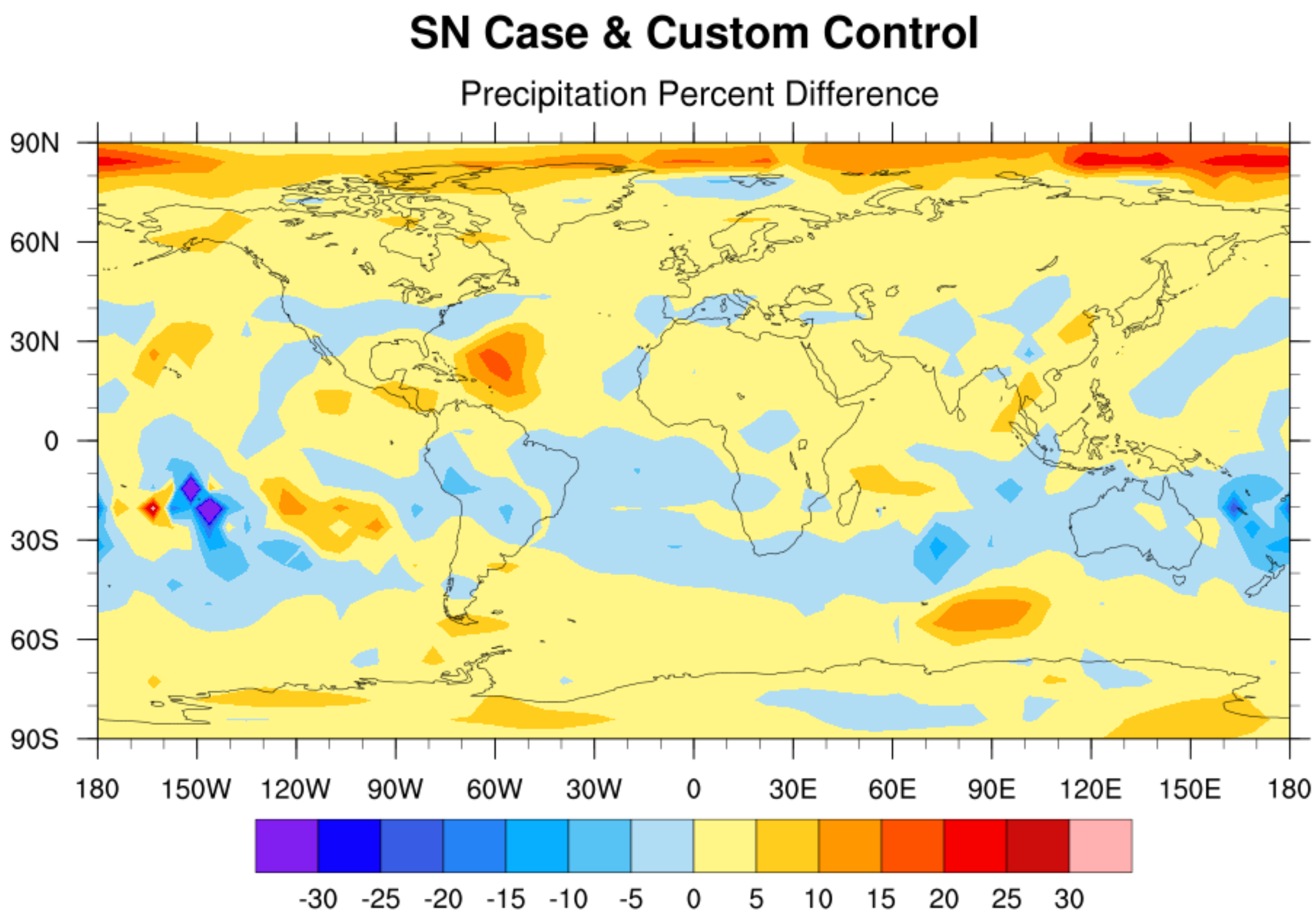

Figure 9: Percent difference in precipitation, averaged over the ensemble and over the last 30 years of the runs. Positive/negative numbers indicate higher/lower precipitation values in the SN case.

## VII. Conclusions

Past work investigating the impact of astrophysical ionizing radiation events, including supernovae [9,12,13,33–35], gamma-ray bursts [10,11,36–42], and extreme solar storms [21,22,43], has primarily focused on depletion of stratospheric ozone and the effects of subsequent increase in solar UV irradiance on the ground and in the upper ocean. Some consideration has been given to other effects such as increased nitrate deposition [44,45], damaging ground-level ozone [46], increased lightning and wildfire [8], and the possibility of climate change following a drawdown of $CO_2$ [47]. Considerations of potential climate change has been limited to broad estimates or simplified calculations [15,47,48].

Initial estimates of the climate impact of a supernova [15] assumed that $O_3$ depletion would act as a reduction of greenhouse forcing, leading to global cooling. Similarly, increased $NO_2$ concentration due to astrophysical ionizing radiation was initially proposed as a cooling mechanism, due to its absorption band in the short-wavelength visible part of the spectrum [10,48]; however, more detailed work has shown that the removal of $O_3$ offsets this due to its own absorption band in visible wavelengths, leading instead to a slight increase in visible light irradiance at the ground [39].

Ours is the first study to employ a climate model to investigate the question of how $O_3$ depletion following a supernova would affect Earth's climate. We have used a specific SN case motivated by geochemical evidence, which delivers an increase in cosmic radiation over 100s to 1000s of years. Broadly speaking, the climate effects at a global scale appear minor. Contrary to early work that anticipated global cooling, we find an increase in globally averaged surface temperature; however, the change is quite small. It is worth noting that this is the opposite of what was found in two studies using the PlaSim model that examined the change after removing $O_3$ entirely, in which a decrease in globally averaged surface temperature of several degrees was observed [25,26]. That difference can be attributed to the fact that removing $O_3$ entirely reduces the overall greenhouse gas forcing, while in our case $O_3$ was only reduced at high altitudes.

While the globally averaged change in temperature is small, regional temperature changes are larger in some areas, especially in the Northern polar region. In addition, we find shifts in atmospheric transport and precipitation patterns. Any change in temperature, winds, and precipitation will have consequences for local ecosystems. The details of such impacts are beyond the scope of this work, but our study illustrates that there is a range of possible effects due to changes in atmospheric $O_3$ concentrations, beyond the usual issue of increased solar UV exposure.

While motivated by geochemical evidence for one or more nearby supernovae about 2.6 million years ago, our study was not designed as a paleoclimate investigation and therefore cannot be applied directly to that case. While the global effect we have found is small, a model setup that is carefully tuned to simulate the appropriate paleoclimate conditions might be able to identify geographical patterns, such as precipitation changes, that could be linked to the geologic and fossil records. More broadly, our study motivates further exploration of the climate effects of other astrophysical ionizing radiation events such as gamma-ray bursts.

Acknowledgements

The authors are very grateful to Frank Lunkeit for assistance with using PlaSim, including help with integrating prescribed $O_3$ profiles. Analysis and plotting was done using the NCAR Command Language (Version 6.6.2, Boulder, Colorado: UCAR/NCAR/CISL/TDD. http://dx.doi.org/10.5065/D6WD3XH5).